\documentclass[aps, prl, reprint, groupedaddress, superscriptaddress, amsmath, amssymb]{revtex4-1} 
\usepackage[colorlinks,allcolors=blue]{hyperref}
\usepackage{epstopdf}
\usepackage{graphicx}
\usepackage{braket}
\usepackage{xcolor}

\begin{document}


\title{Integrated Plasmonics: Broadband Dirac Plasmons in Borophene}

\author{Chao Lian}
\affiliation{Beijing National Laboratory for Condensed Matter
	Physics and Institute of Physics, Chinese Academy of Sciences,
	Beijing, 100190, China}
\author{Shi-Qi Hu}
\author{Jin Zhang}
\affiliation{Beijing National Laboratory for Condensed Matter
    Physics and Institute of Physics, Chinese Academy of Sciences,
    Beijing, 100190, China}
\affiliation{University of Chinese Academy of Sciences, Beijing, 100049, P. R. China}
\author{Cai Cheng}
\affiliation{Beijing National Laboratory for Condensed Matter
	Physics and Institute of Physics, Chinese Academy of Sciences,
	Beijing, 100190, China}
\author{Zhe Yuan}
\affiliation{Center for Advanced Quantum Studies and Department of Physics, Beijing Normal University, Beijing 100875, China}
\author{Shiwu Gao}
\email{swgao@csrc.ac.cn}
\affiliation{Beijing Computational Science Research Center, Beijing, 100193, China}
\author{Sheng Meng}
\email{smeng@iphy.ac.cn}
\affiliation{Beijing National Laboratory for Condensed Matter
    Physics and Institute of Physics, Chinese Academy of Sciences,
    Beijing, 100190, China}
\affiliation{University of Chinese Academy of Sciences, Beijing, 100049, P. R. China}
\date{\today}

\begin{abstract}
   The past decade has witnessed numerous discoveries of two-dimensional (2D) semimetals and insulators, whereas 2D metals are rarely identified. Borophene, a monolayer boron sheet, has recently emerged as a perfect 2D metal with unique structure and electronic properties. Here we study collective excitations in borophene, which exhibit two major plasmon modes with low damping rates extending from infrared to ultraviolet regime. The anisotropic 1D plasmon originates from electronic excitations of tilted Dirac cones in borophene, analogous to that in heavily doped Dirac semimetals. These features make borophene promising to realize directional polariton transportation and broadband optical communications for next-generation optoelectronic devices.
\end{abstract}
\maketitle

When propagating along the metal-dielectric interface in plasmonic devices, electromagnetic waves couple with electronic motions and form surface plasmon polaritons (SPPs)~\cite{Pitarke2006, Maier2007a}. Noble metal films (e.g. Ag and Au) provide abundant free electrons to generate high-frequency plasmons in SPP devices~\cite{Gao2011a, Thongrattanasiri2012, Sessi2015, Zubizarreta2017}. However, the SPPs in these devices suffer from low confinements and significant losses during propagation~\cite{sundararaman_plasmonics_2020, Slotman2018, scienceconfined}, resulted from the manifold interband damping and strong plasmon-phonon scatterings~\cite{West2010, Jablan2009}.

Naturally, ultrathin two-dimensional (2D) materials, such as graphene~\cite{ACSnano_graphene2014, ACSphton_graphene2014, gradoped2018, graphene_noss_nature2018},  phosphorene~\cite{Low2014a, Ghosh2017a, PhysRevB.96.115402, PhysRevB.92.085406, Edo2018}, and MoS$_2$~\cite{Scholz2013, Liu2015c, Yue2017, Sen2017, doi:10.1021/acs.nanolett.5b04588, SMLL:SMLL201403422,mos2jpcm2017},
are proposed to generate SPPs with low damping rates and high confinements due to stronger light-matter interactions~\cite{DasSarma1981, Hwang2007b, Yuan2008, Dai2015a, Cox2015, Fei2012, PhysRevB.77.081411, Chen2012, Ansell2014, DeAbajo2014, Xia2014, Thakur2016a, fang_nanoplasmonic_2015, VacacelaGomez2016, Marusic2017,Yan2011, Despoja2018, hu_gas_2019}. However, low carrier densities in these materials limit the frequencies of the plasmonic response up to terahertz or infrared region, where light sources and optoelectronic detectors are less developed~\cite{Gao2011, Koppens2011, Low2014, Low2016, doi:10.1021/nl400601c, doi:10.1021/nl501096s, Ni2016}. The 2D materials with higher carrier density and higher plasmon frequencies are particularly desirable for building optical devices with beyond-diffraction-limit resolutions, detecting biotechnological processes, and enhancing atomic transitions~\cite{Davy2015, Costas2011, Pendry2000}. 

Recently, borophene, a monolayer boron sheet, has been experimentally synthesized either on a solid substrate via molecular beam epitaxy~\cite{Mannix2015a, Feng2016c} or as free-standing atomic sheets via sonochemical liquid-phase exfoliation~\cite{Vinu2019}. Borophene has extraordinary electric, optical and transport properties which are highly related to its intrinsic metallic, Dirac-type band structures~\cite{Feng2017a, Feng2018, Zhang2018,Feng2016, Zhao2016c, Penev2016, Jalali2018, Yuefei2017}. The density of the Dirac electrons in borophene is extremely high ($10^{15}$~$e$/cm$^{2}$)~\cite{Feng2017a,Feng2018,Zhang2018} compared to doped graphene ($10^{12}\sim10^{14}$~$e$/cm$^{2}$)~\cite{PhysRevLett.105.256805}. Thus, we expect that borophene, as an intrinsic 2D metal with both high carrier densities and high confinements, can be a promising candidate to build low-loss broadband SPP devices.

\begin{figure*}
    \centering
    \includegraphics[width=0.8\linewidth]{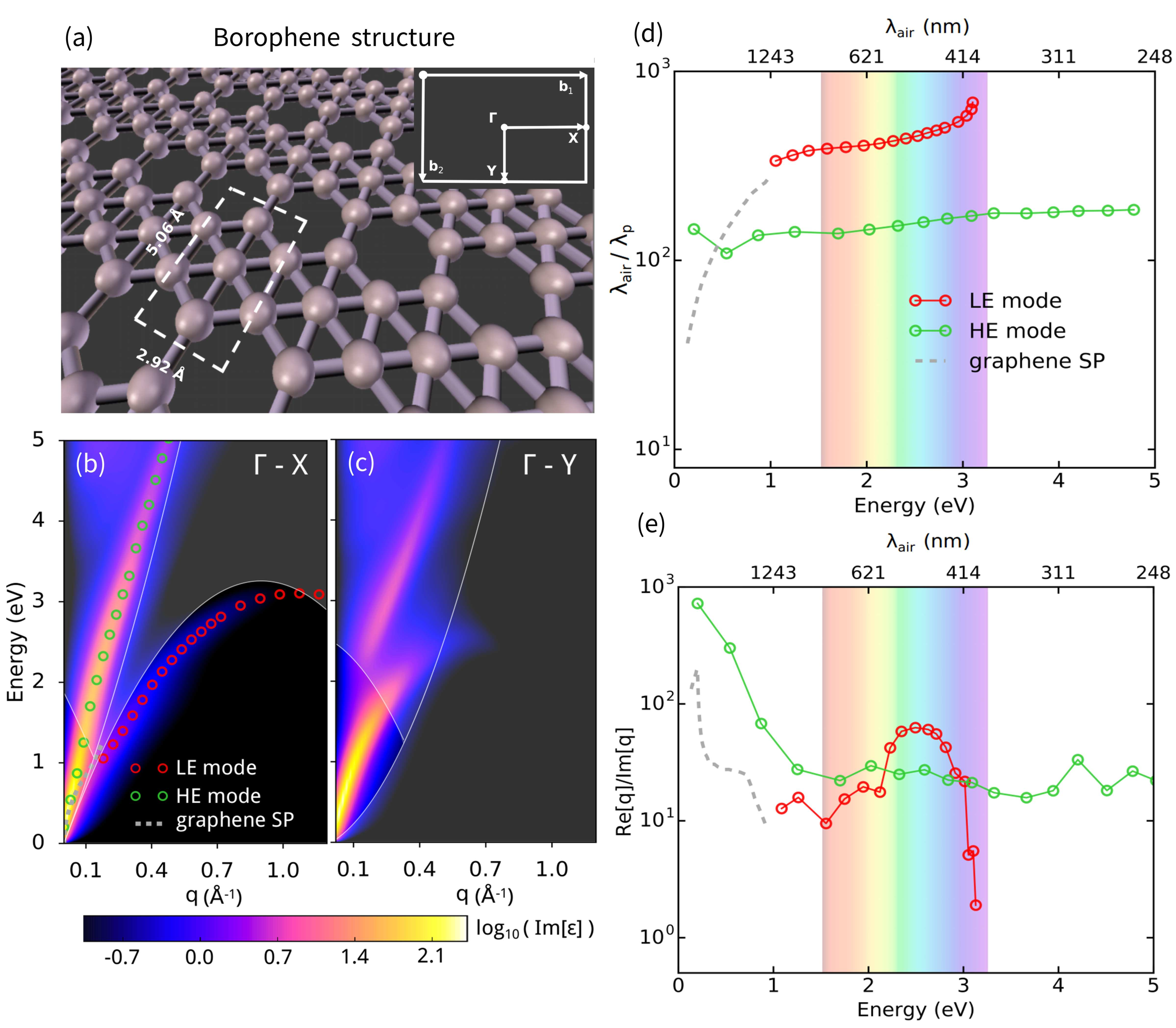}
    \caption{(a) Structure and Brillouin zone of $\beta_{12}$ borophene. (b-c) Imaginary parts of dielectric functions along the $\Gamma$-X and $\Gamma$-Y directions. Shaded areas denote the single-particle excitations (SPE) regions. Green and red circles denote the peaks of the high-energy (HE) and low-energy (LE) mode, respectively. The dashed line in (b) represents the dispersion of surface plasmon in graphene~\cite{Jablan2009}. (d) Confinement ratio $\lambda_{air}/\lambda_{p}$ and (e) loss $\mathrm{Re}[q]/\mathrm{Im}[q]$ of the borophene plasmon as a function of wavelength.
    The carrier density in graphene is $n=3\times10^{13}$~cm$^{-2}$. The colored areas denote the energy range of the visible light with the corresponding color.} 
    \label{fig:Figure1}
\end{figure*}

In this work, we report discovery of low-loss and highly-confined broadband plasmons in borophene, based on time-dependent density functional theory (TDDFT)~\footnote{The frequency and wavevector dependent density response functions are calculated within the TDDFT formalism using random phase approximation as implemented in the \textsc{\scriptsize GPAW} package~\cite{gpaw1,gpaw2,gpawlr,ase1,ase2}. The projector augmented-waves method and Perdew-Burke-Ernzerhof exchange-correlation~\cite{Perdew1996} are used for the ground state calculations. The plane-wave cutoff energy is set to be 500~eV. The thickness of vacuum layer is set to be larger than 10 {\AA}. The Brillouin zone is sampled using the Monkhorst-Pack scheme~\cite{monkhorst1976special} with a dense \textbf{k}-point mesh $142\times72\times1$ in the self-consistent calculations.}. In our calculations, we observe two plasmon branches: A high energy (HE) mode extends to ultraviolet and originates from collective excitations of bulk electrons in the 2D material; in the low energy (LE) region, a new plasmon mode exhibits a strong anisotropic behavior and broadband response. The new plasmon mode originates from collective electronic transitions of one-dimensional (1D) electron gas derived from tilted Dirac cones. Both modes show remarkable low-loss properties comparable to graphene, but at significant higher frequencies, thanks to borophene's high carrier density and low-dimensional nature~\cite{Jablan2009}. The confinement of plasmon in borophene is also 2-3 orders of magnitude higher than that in Ag~\cite{2DAg2018}. The discovery of novel plasmon modes make borophene more suitable than graphene and noble metals for plasmon generation and integrated optoelectronics working at a broad range of frequencies.

Figure~\ref{fig:Figure1} shows the atomic structure of the $\beta_{12}$ borophene. The $\beta_{12}$ borophene is the most stable phase found in experiments~\cite{Mannix2015a, Feng2016c,Vinu2019} and is thus chosen as a representative structure of borophene. The unit cell is rectangular with the lattice parameters $a_1=2.92$~\AA\ and $a_2=5.06$~\AA, consisting of five boron atoms. Periodic vacancies line up along the horizontal direction (denoted as the X direction). This special structure introduces anisotropy between the horizontal (X) and vertical (Y) directions in both the real and reciprocal space [Fig.~\ref{fig:Figure1}(a)].

\begin{figure*}
	\centering
	\includegraphics[width=1.0\linewidth]{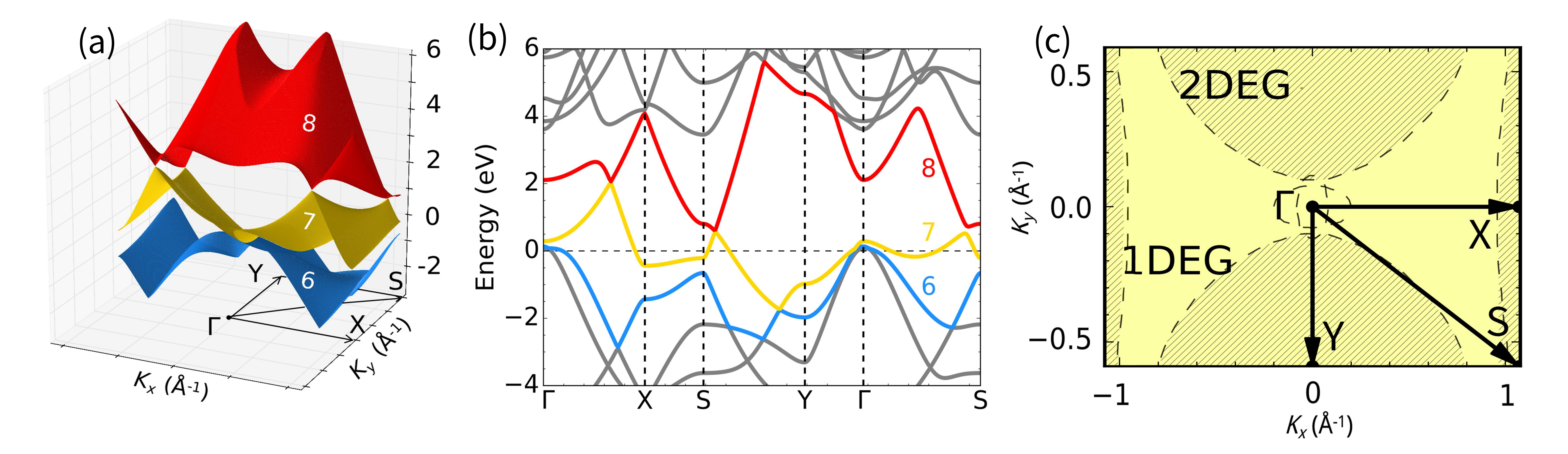}
	\caption{(a) Three-dimensional band structures, (b) two-dimensional band structures, and (c) Fermi surface of $\beta_{12}$ borophene.}
	\label{fig:Figure2}
\end{figure*}

The optical absorption spectra, obtained from the imaginary parts of the dielectric functions, are shown in Fig.~\ref{fig:Figure1}(b-c). We observe two plasmon branches in different broad energy ranges in our calculations. A high energy branch (HE mode) disperses almost linearly and extends to the ultraviolet regime ($>5.0$~eV) at the momentum range $q>0.4$~\AA$^{-1}$. This mode is almost isotropic along the X and Y directions. In contrast, a low energy branch (LE mode) shows evidently anisotropic dispersions along different directions. Along the $\Gamma$-X direction, the LE branch shows an inverted parabolic dispersion over the whole Brillouin zone, with the maximum energy at the half of the reciprocal lattice vector, $q = |\mathbf{b_1}|/2 = 1.07$~\AA$^{-1}$. Along the $\Gamma$-Y, however, only the HE mode shows up at small $q$ regimes; the LE mode develops only at energies higher than $\sim$2 eV, whereas the two plasmon modes strongly hybridize with each other. The features are significantly different from the behaviors along the $\Gamma$-X direction.  

Both the HE and LE branches can form a broadband SPP with low losses. As shown in Fig.~\ref{fig:Figure1}(d) and (e), we calculate the confinement ratio $\lambda_{air}/\lambda_{p}$ and relative loss $\mathrm{Re}[q]/\mathrm{Im}[q]$ of borophene plasmons, where $\lambda_{air}$ and $\lambda_{p}$ are the light wavelengths in air and borophene, respectively~\cite{Jablan2009}. The LE plasmons possess high confinement ratios $\lambda_{air}/\lambda_{p}$ of 330--700 and long losses $\mathrm{Re}[q]/\mathrm{Im}[q]$ of 10--700 at different wavelengths from 400 to 1240~nm. The $\lambda_{air}/\lambda_{p}$ and $\mathrm{Re}[q]/\mathrm{Im}[q]$ of the HE plasmon are slightly lower over a broader energy range. Both the confinement ratios and losses are comparable to those in heavily doped graphene ($\lambda_{air}/\lambda_{p}\sim300$ and $\mathrm{Re}[q]/\mathrm{Im}[q] = 10-200$)~\cite{Jablan2009} and much larger than those in the Al or Ag films ($\lambda_{air}/\lambda_{p} \approx1$)~\cite{2DAg2018}. Furthermore, the low damping SPPs only exists within the infrared regime ($\lambda_{air}>1240$~nm) in graphene~\cite{Jablan2009, gradoped2018, mos2jpcm2017, Ghosh2017a}, while borophene can generate the low-loss SPPs in a much broader energy range from infrared to ultraviolet. This indicates that borophene is a better building material for the low-loss broadband optoelectronic devices. 

\begin{figure*}
	\centering
	\includegraphics[width=1.0\linewidth]{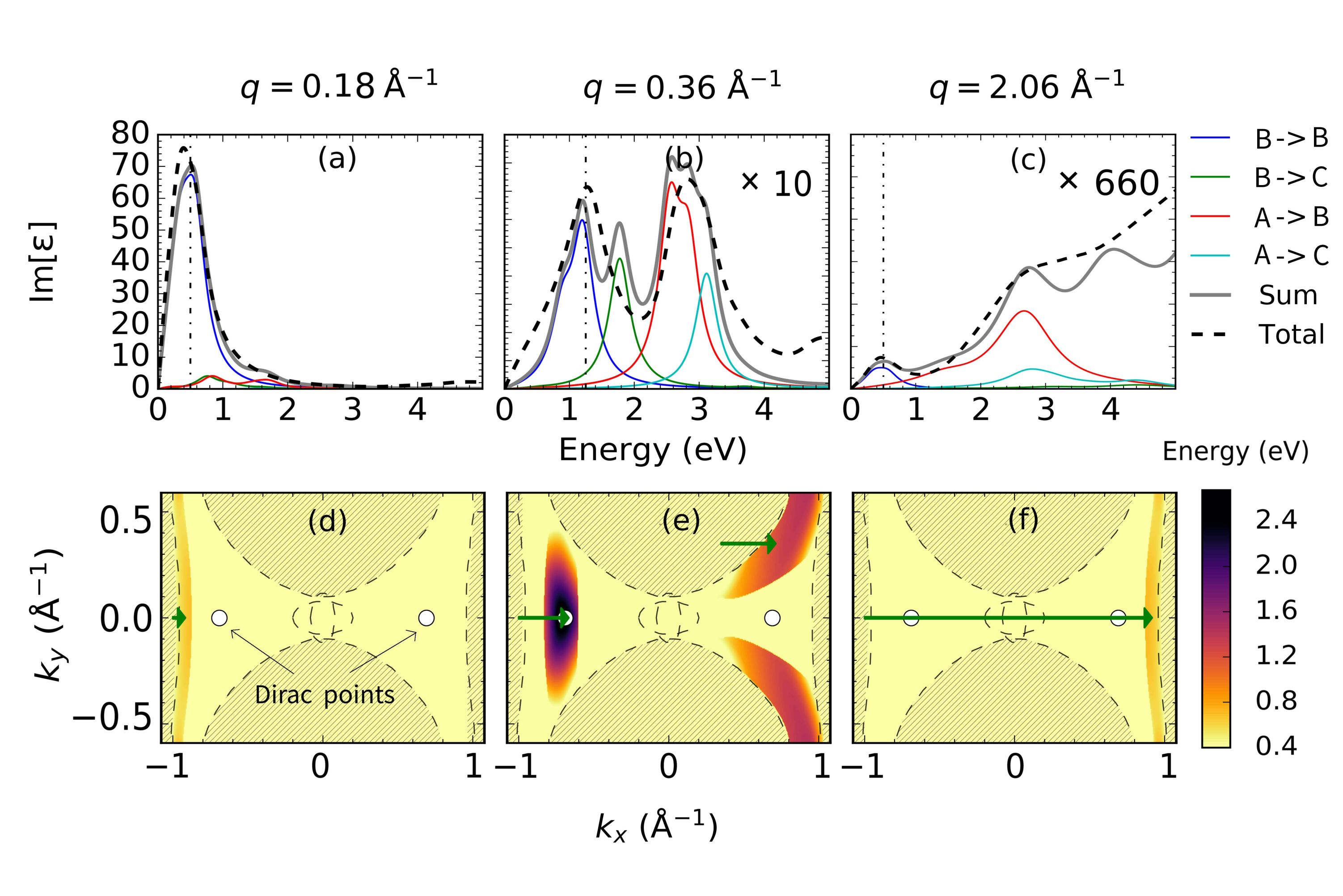}
	\caption{(a-c) Imaginary part of the dielectric function at different momenta $q$. Transition $X \to Y$ denotes the component contributed by the excitation from band X to Y ($X, Y \in \{A, B, C\}$) [Fig.~\ref{fig:Figure2}(a) and (b)]. The grey dashed line denotes the sum of contributions by the excitation from band X to Y ($X, Y \in \{A, B, C\}$) for (a) and (b), while for (c) excitation between $\{A, B, C\}$ and 5 higher bands are also included. (d-f) The contour plot of the energy difference of the $B \to B$ intraband transition $\omega_{B,B}(q,k) = \epsilon_{B,k} - \epsilon_{B,k-q}$ at different $q$. The green arrows denote the excitation from $k-q$ to $k$. The gray dashed lines denote the Fermi surface.
	}
	\label{fig:Figure3}
\end{figure*}

\begin{figure}
	\centering
	\includegraphics[width=1.0\linewidth]{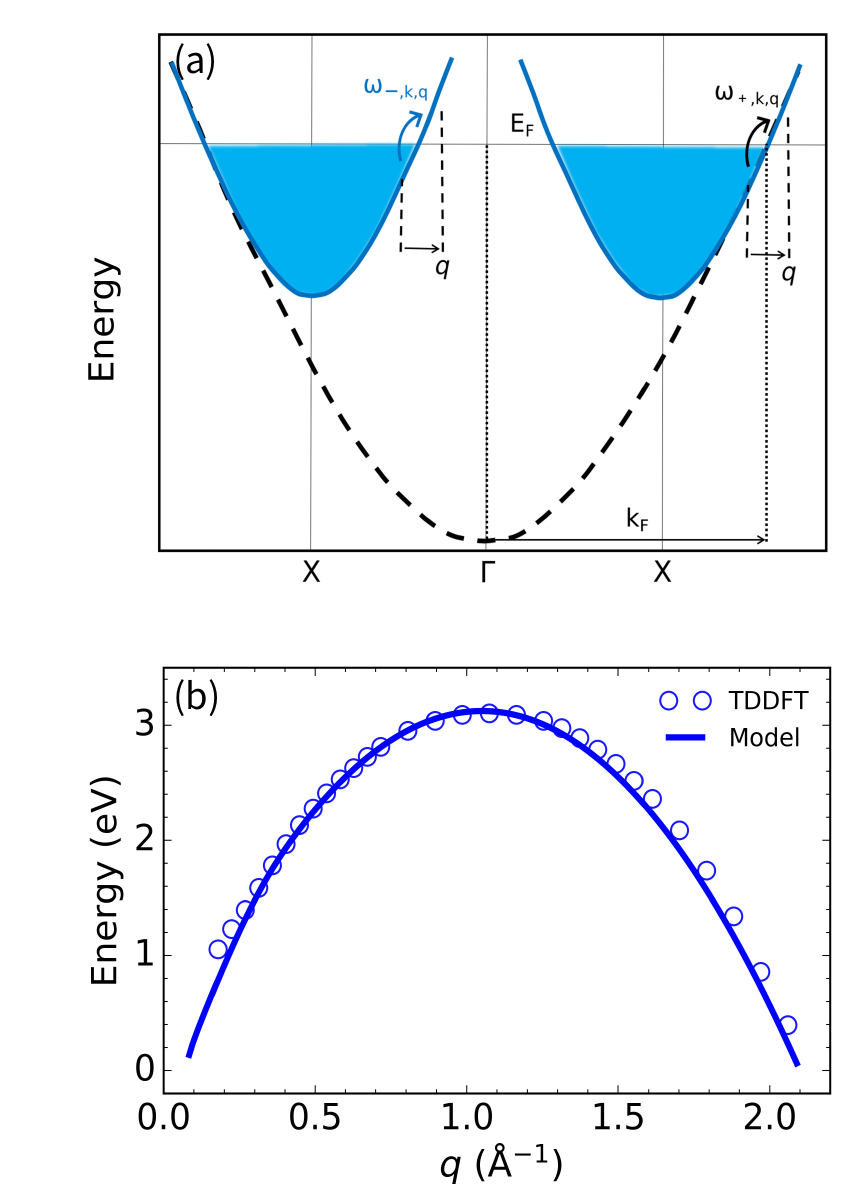}
	\caption{(a) Schematic band structure along X-$\Gamma$-X. The dash line follows $E_k = k^2/2m^* +E_0 $ with $E_0=-3.26$~eV and  $m^* = 1.6 m_0$, representing the band of 1DEG@$\Gamma$ with the same dispersion as that of 1DEG@X. The black and blue arrows denote the excitation channel of 1DEG@$\Gamma$ and 1DEG@$X$, respectively. (b) The LE plasmon dispersions calculated by the 1DEG model (solid line) and TDDFT (open circles).}
	\label{fig:Figure4}
\end{figure}

In the following paragraphs, we discuss the electronic origin of these two outstanding characteristics, the anisotropy and low-loss rate, of the borophene plasmon. We first note that the anisotropic plasmon in borophene is \textit{not} a trivial consequence of the rectangular lattice, which generates only a weak anisotropy in phosphorene plasmons~\cite{Ghosh2017a}. Instead, the unique electronic structure is the major reason of the anisotropy. As shown in Fig.~\ref{fig:Figure2}(a) and (b), band B crosses the Fermi energy and joins band C at the Dirac points at 2~eV, indicating the metallic nature of borophene and forming the Fermi surface, as shown in Figure~\ref{fig:Figure2}(c). Another Dirac point forms along the S--Y direction and at a lower energy, 0.5 eV. Both Dirac cones are tilted in their shape, consistent with experimental measurements~\cite{Feng2017a,Feng2018}. Thus, borophene forms a Fermi surface comprising three parts: (I) a ribbon along $\Gamma$-Y centered at X, implying a 1D electron gas (1DEG) from tilted Dirac electrons subjecting to strong confinement along the Y direction; (II) two semicircular regions characterizing a bulk 2D electron gas;  (III) a small hole pocket near the $\Gamma$ point.

We ascribe the anisotropic LE mode to the intraband oscillations of the 1DEG between the Dirac cones. As shown in Fig.~\ref{fig:Figure3}, we analyze the contributions of different electron-hole transitions to the plasmonic peaks within the independent particle approximation (IPA). As shown in Fig.~\ref{fig:Figure3}(a-c), the HE mode emerges mainly at $\sim3.8$~eV for $q=0.36$~\AA$^{-1}$, which comes from mixed interband transitions of band A $\to$ B, band A $\to$ C and band B $\to$ C. In comparison, the LE peaks are located at 1.08, 1.79, and 0.5~eV for $q=0.18$, 0.36, and 2.06~\AA$^{-1}$, respectively, which are dominated by the intraband transitions B $\to$ B. Accordingly, we visualize the excitation mode of the LE plasmon in the reciprocal space. The contour plot of the energy difference, $\omega_{B,B}(q,\mathbf{k})$, between the initial state $\{B, \mathbf{k}- \mathbf{q}\}$ and final state $\{B,\mathbf{k} \}$, where $ \omega_{B,B}(q,\mathbf{k}) = \epsilon_{B\mathbf{k}} - \epsilon_{B \mathbf{k} - \mathbf{q}}$ for $\epsilon_{B\mathbf{k}}>0 \bigcap \epsilon_{B \mathbf{k} - \mathbf{q}} < 0$, $\epsilon_{B\mathbf{k}}$ is the eigenvalue of band B at the $\mathbf{k}$ point. As shown in Fig.~\ref{fig:Figure3}(d-f), the oscillation of the 1DEG dominates the LE mode and leads to its anisotropy, since the 1DEG can only oscillate along $\Gamma$-X direction. Specifically, at $q=0.36$~\AA$^{-1}$, the LE plasmon is generated by electron excitations from the Fermi surface to the Dirac points~\cite{Feng2017a}. Thus, borophene can be viewed as an extremely hole-doped graphene in generating the LE plasmon.

To explain the mechanism of the low-damping characteristic, we adopt the confined 1DEG model~\cite{DasSarma1996, Gao2005, Yan2007b} that is widely used to describe plasmons of atomic chains~\cite{Ritsko1975, Rugeramigabo2010, Nagao2006, Yan2008, Liu2008b, Liu2008a}. We find an additional excitation channel exists in the borophene for its special 1DEG centered at X point (1DEG@X), compared with the usual 1DEG@$\Gamma$: For excitations at certain momentum $q$, the conventional channel for 1DEG@$\Gamma$ [the black arrow in Fig.~\ref{fig:Figure4}(a)] is $\omega_{+,k,q} = \omega_{k+q} - \omega_{k} = (kq + \frac{1}{2}q^2)/m$, with $\omega_k = k^2/2m$ (assuming $\hbar=1$ and $m$ is the effective mass of electron). For 1DEG@X, there is an additional channel (the blue arrow) located on the opposite side ($-k_F$) $\omega_{-,k,q} = \omega_{-k+q}-\omega_{-k} =(-kq + \frac{1}{2}q^2)/m$. 

Accordingly, we calculate the plasmon dispersions of these two excitation channels, which are determined from the zeros of the dielectric function $\label{pl_cond} \epsilon(q,\omega_p) = 1 - V(q) \mathrm{Re}\chi_0(q,\omega_p) = 0$. Here, $\omega_p$ is the frequency of plasmom, $V(q)$ is the Coulomb potential, $\chi_0$ is the response function with IPA~\cite{Friesen1980}
$\label{chi0_1}
\chi_0(q,\omega_p) = \frac{1}{\pi} \int dk \theta(q \pm k-k_F)  \theta(k_F \mp  k) 		
\left[ \frac{1}{\omega - \omega_{\pm,k,q} + i\eta} - \frac{1}{\omega  + \omega_{\pm,k,q} + i\eta} \right]
$. Thus, $\label{ReChi0_1} \mathrm{Re}\chi_0(q,\omega_p) = \pm \frac{1}{\pi q}\ln \frac{\omega^2 - \omega^2_- }{\omega^2 - \omega^2_+}$, 
where $\label{eq:pair-exc} \omega^2_\pm = \left| k_F q \pm \frac{1}{2} q^2\right|/m$ are the upper (+) and lower (--) limits of the single-particle excitation (SPE) regimes, respectively. The 1D plasmon dispersions can be solved as
$\label{pl_1D_1} \omega^+_p(q) = \sqrt{\left[A(q)\omega^2_+  - \omega^2_- \right]/\left[A(q) - 1\right]}$ and $\label{pl_1D_2} \omega^-_p(q) = \sqrt{\left[A(q)\omega^2_- - \omega^2_+ \right]/\left[A(q) - 1\right]}$~\cite{DasSarma1996,Friesen1980},
where $\label{pl_act} A(q) = \exp\left\{ { \pi q }/\left[mV(q)\right]\right\}$, $V(q) = {e^2}/{K} e^{b^2 q^2}\int_{b^2q^2}^{\infty} {e^{-x}}/{x} dx$, $K$ is the static dielectric constant, $b$ is the width of the 1DEG, and $m$ is the effective mass of electrons the 1DEG.

As shown in Fig.~\ref{fig:Figure4}(b), while $\omega^+_p(q)$ hybridizes and merges with the isotropic HE mode, $\omega^-_p(q)$ accurately reproduces the inverse parabolic dispersion calculated from TDDFT, with the parameters $k_F =1.07$~\AA$^{-1}$, $b=$4~\AA, $K=6.25$ and $m = 2.4 m_0$. We note that, among the parameters, $k_F$ is the length of $\Gamma$-X and thus not adjustable; 1DEG width $b$ and dielectric constant $K$ have negligible effects on $\omega_p(q)$ when $q > 0.1$~\AA$^{-1}$. The effective mass $m = 2.4 m_0$ is slightly larger than that evaluated from the band structure $m^* = 1.6 m_0$, due to many body screening effects. This indicates that our model directly and robustly reflects the electronic origin of novel borophene plasmon.

The same model may also explain the LE plamon branch along the $\Gamma$-Y direction. Along $\Gamma$-Y, electrons first undergo an interband excitation to energy levels at $\sim$2 eV above the Fermi level, where electrons form a similar 1DEG confined along $\Gamma$-Y direction. With strong interband transitions included in plasmon excitations, excitations of the 1DEG at $\Gamma$-Y have similar excitation channels as 1DEG along $\Gamma$-X discussed above. Therefore, a kink in plasmon dispersion at $\sim$2 eV is formed, followed by an inverse parabolic dispersion along the $\Gamma$-Y direction.

Based on these results, we discuss the mechanism of the low-damping behavior of the LE plasmon. With $q \rightarrow \infty$, these two branches $\omega^-_p(q)$ and $\omega^+_p(q)$ yield different asymptotic behaviors. The $A(q)$ dominates when $q \rightarrow \infty$ and leads $\omega^+_p(q) \to \omega_{+}$ and $\omega^-_p(q) \to \omega_{-}$. That is, $\omega^+_p(q)$ [$\omega^-_p(q)$] is always higher (lower) than the SPE region. 
Moreover, the Dirac-type intraband excitations [Fig.~\ref{fig:Figure3}(e)] further suppress the SPE region, since the pseudospin symmetry forbids perpendicularly-polarized excitations~\cite{Hwang2007b}. The combined effects remarkably produce the low-loss LE plasmons.


We briefly discuss the technical requirements of building realistic plasmonic devices with borophene. The first important requirement is the stability of borophene under ambient conditions. Feng et al. showed that borophene on the Ag substrate is generally robust to the oxidation~\cite{Feng2016c}. Recently, Ranjan et al. synthesized large-scale free-standing borophene by liquid-phase exfoliation~\cite{Ranjan2019}. They claimed that the free-standing borophene layer is even more stable against oxidation. This indicates that a borophene-based device is applicable under ambient conditions.

Other than the stability, the plasmonic application requires large-scale high-quality borophene. The point defect in borophene barely affects its band structures [Fig.~\ref{fig:figureadd2}]. Furthermore, the charge doping has little effect on the plasmonic response in borophene, as shown in Fig.~\ref{fig:DopingEELScontour}. Thus, the borophene plasmon is insensitive to a small number of point defects, similar to the experimental observation in graphene~\cite{PhysRevB.84.033401, Langer_2010}. However, the anisotropic plasmon will probably be substantially smeared in polycrystalline borophene. Synthesis of centimeter-scale single-crystalline borophene is needed for the plasmonic device, which is quite promising considering the achievements in the growth of wafer-scale graphene~\cite{Lee2010GNGrowth, Lin2011GNGrowth}, transition metal dichalcogenides~\cite{Liu2020GrowthRev}, and boron nitride~\cite{Wang2019BN}. 
We hope that the unique plasmon properties predicted in our work could motivate the technical progress in preparing a high-quality borophene sample for building the potential plasmonic devices.

Absent in other 2D materials, we observe in our calculations the coexistence and interplay of the 1DEG, 2DEG and Dirac electrons in the collective plasmon excitations in metallic borophene.
The exotic features such as low-loss, strong confinement and panchromatic responses of LE and HE plasmons 
make borophene a promising candidate for applications in nanophotonics and integrated optoelectronics working at broadband frequencies.\\

We acknowledge insightful discussions with Prof. Ling Lu. This work is partially supported by MOST (grants 2016YFA0300902 and 2015CB921001), NSFC (grant 11774396 and 91850120), and CAS (XDB07030100). S.G. acknowledges supports from MOST through grants 2017YFA0303404 and 2016YFB0700701, NSFC through grant NSAF U-1530401.

C.L. and S.H. contribute equally to this work. 

%


\clearpage
\widetext
\setcounter{table}{0}
\renewcommand{\thetable}{S\arabic{table}}%
\setcounter{figure}{0}
\renewcommand{\thefigure}{S\arabic{figure}}%
\setcounter{equation}{0}
\renewcommand{\theequation}{S\arabic{equation}}%
\section{Supplemental Material}

\noindent \textbf{Methods}\\
\textbf{Linear response theory for plasmon excitations.}
The frequency and wave-vector dependent density response functions are calculated within time-dependent density-functional theory (TDDFT) formalism using the random phase approximation for exchange-correlation functional. 
The non-interacting density response function in real space is written as
\begin{equation}
\begin{split}
\chi^0(\mathbf{r}, \mathbf{r}^{\prime}, \omega) =& \sum_{\mathbf{k}, \mathbf{q}}^{\mathrm{BZ}}\sum_{n, n^{\prime}}  
\frac{f_{n\mathbf{k}}-f_{n^{\prime} \mathbf{k} + \mathbf{q}}}{\omega + \epsilon_{n\mathbf{k}} - \epsilon_{n^{\prime} \mathbf{k} + \mathbf{q}} + i\eta} \times \\ 
& \psi_{n\mathbf{k}}^{\ast}(\mathbf{r}) \psi_{n^{\prime} \mathbf{k} + \mathbf{q} }(\mathbf{r}) \psi_{n\mathbf{k}}(\mathbf{r}^{\prime}) \psi^{\ast}_{n^{\prime} \mathbf{k} + \mathbf{q} }(\mathbf{r}^{\prime}),
\end{split}
\end{equation} 
where $\epsilon_{n \mathbf{k}}$ and $\psi_{n \mathbf{k}}(\mathbf{r})$ are the eigenvalues and eigenvectors of the ground state Hamiltonian. For translation invariant systems,  $\chi^0$ can be expanded in planewave basis as
\begin{equation}
\chi^0(\mathbf{r}, \mathbf{r}^{\prime},  \omega) = \frac{1}{\Omega} 
\sum_{\mathbf{q}}^{\mathrm{BZ}} \sum_{\mathbf{G} \mathbf{G}^{\prime}}
e^{i(\mathbf{q} + \mathbf{G}) \cdot \mathbf{r}} \chi^0_{\mathbf{G} \mathbf{G}^{\prime}}(\mathbf{q}, \omega) 
e^{-i(\mathbf{q} + \mathbf{G}^{\prime}) \cdot \mathbf{r}^{\prime}} ,
\end{equation} 
where $\Omega$ is the normalization volume, $\mathbf q$ stands for the Bloch vector of the incident wave and $\mathbf G (\mathbf G^{\prime})$
are reciprocal lattice vectors.
The full interacting density response function is obtained by solving the Dyson's equation, from its non-interacting counterpart $\chi^0$ as
\begin{equation}
\begin{split}
\chi(\mathbf r, \mathbf{r^{\prime}}, \omega) =& \chi_0(\mathbf r,  \mathbf{r^{\prime}}, \omega) \\ 
+& \iint_{\Omega} d\mathbf{r}_1 d\mathbf{r}_2 \chi_0(\mathbf r, \mathbf{r}_1, \omega)
K(\mathbf{r}_1, \mathbf{r}_2) \chi(\mathbf{r}_2,  \mathbf{r^{\prime}} ,\omega),
\end{split}
\end{equation}
where the kernel is the summation of coulomb and exchange-correlation (XC) interaction
\begin{equation}
K(\mathbf{r}_1, \mathbf{r}_2) = \frac{1}{|\mathbf{r}_1 -\mathbf{r}_2|} 
+ f_{xc}.
\end{equation}
Here, $f_{xc} = {\partial V_{xc}[n]}/{\partial n}$ is the XC kernel. 
\newcommand{\RefcComd}{{The common used XC kernels include adiabatic local density approximation (ALDA)~\cite{Perdew1981}, Bootstrap approximation~\cite{PhysRevLett.107.186401}, Q-dependent kernals~\cite{giuliani2005quantum, corradini_analytical_1998}, etc. 
A simplest case is the so-called random phase approximation (RPA), with $f_{xc} = 0$. Since the plasmon is demonstrated to be well described in RPA~\cite{Marques}, we use RPA in the following calculation, while some results with other XC kernels such as ALDA are also tested and found to be consistent with RPA results. As indicated in Ref.~\cite{cazzaniga_dynamical_2011}, many-body local field effect is not a dominant factor in simple metals such as sodium and aluminum. Thus, although local field effect may be an interesting topic, the Q-dependent kernels are not tested and discussed in this work.}}
{\label{txt:RefcComd} \RefcComd}

With translational symmetry, it is more convenient to represent $\chi^0$ in the reciprocal lattice space. Fourier coefficients $\chi^0_{\mathbf{G} \mathbf{G}^{\prime}}(\mathbf{q}, \omega)$ are written as
\begin{equation}
\begin{split}
\chi^0_{\mathbf{G} \mathbf{G}^{\prime}}(\mathbf{q}, \omega) = & 
\sum_{n, n^{\prime}} \chi^0_{\mathbf{G} \mathbf{G}^{\prime} n, n^{\prime} }(\mathbf{q}, \omega)
\end{split}
\end{equation}
where
\begin{equation}
\label{chi0nn'}
\begin{split}
\chi^0_{\mathbf{G} \mathbf{G}^{\prime} n, n^{\prime} }(\mathbf{q}, \omega) = & \frac{1}{\Omega} 
\sum_{\mathbf{k}}^{\mathrm{BZ}} 
\frac{f_{n\mathbf{k}}-f_{n^{\prime} \mathbf{k} + \mathbf{q} }}{\omega + \epsilon_{n\mathbf{k}} - \epsilon_{n^{\prime} \mathbf{k} + \mathbf{q} } + i\eta} \\ 
&\times \langle \psi_{n \mathbf{k}} | e^{-i(\mathbf{q} + \mathbf{G}) \cdot \mathbf{r}} | \psi_{n^{\prime} \mathbf{k} + \mathbf{q} } \rangle_{\Omega_{\mathrm{cell}}} \\
&\times \langle \psi_{n\mathbf{k}} | e^{i(\mathbf{q} + \mathbf{G}^{\prime}) \cdot \mathbf{r}^{\prime}} | \psi_{n^{\prime} \mathbf{k} + \mathbf{q} } \rangle_{\Omega_{\mathrm{cell}}},
\end{split}
\end{equation}
and so is the the Dyson's equation
\begin{equation}
\label{eq:Dyson}
\begin{split}
\chi_{\mathbf G \mathbf G^{\prime}}(\mathbf q, \omega)  
=& \chi^0_{\mathbf G \mathbf G^{\prime}}(\mathbf q, \omega) \\
+& \sum_{\mathbf G_1 \mathbf G_2} \chi^0_{\mathbf G \mathbf G_1}(\mathbf q,  \omega) K_{\mathbf G_1 \mathbf G_2}(\mathbf q)
\chi_{\mathbf G_2 \mathbf G^{\prime}}(\mathbf q, \omega). 
\end{split}
\end{equation}
The dielectric function can be expressed with $\chi_{\mathbf G \mathbf G^{\prime}}(\mathbf q, \omega)$ as
\begin{equation}
\epsilon^{-1}_{\mathbf G \mathbf G^{\prime}}(\mathbf q, \omega)
= \delta_{\mathbf G \mathbf G^{\prime}} - \sum_{\mathbf{G}_1} K_{\mathbf{G} \mathbf{G}_1}(\mathbf q) \chi_{\mathbf{G}_1 \mathbf G^{\prime}}(\mathbf q, \omega).
\end{equation}
$K_{\mathbf{G} \mathbf{G}_1}(\mathbf q)$ becomes diagonal in reciprocal space
\begin{equation}
K^{\mathrm{Coulomb}}_{\mathbf G \mathbf G_1}(\mathbf q) = 
\frac{4\pi}{|\mathbf q+\mathbf G_1|^2} \delta_{\mathbf G_1 \mathbf G}.
\end{equation}
Note that the kernel $K(\mathbf{q})$ is proportional to $1/q^2$. This indicates that with $q$ increases, the perturbation (the second term in Eq.~\ref{eq:Dyson}) is decreasing and falling into the valid region of lr-TDDFT.

Thus, the dielectric function is simplified as 
\begin{equation}
\epsilon^{-1}_{\mathbf G \mathbf G^{\prime}}(\mathbf q, \omega)
= \delta_{\mathbf G \mathbf G^{\prime}} - \frac{4\pi}{|\mathbf q + \mathbf G|^2} 
\chi_{\mathbf G \mathbf G^{\prime}}(\mathbf q, \omega).
\end{equation}
The macroscopic dielectric function is defined by
\begin{equation}
\label{eq:epsWLFE}
\epsilon_M(\mathbf q, \omega) = \frac{1}{\epsilon^{-1}_{00}(\mathbf q, \omega)}.
\end{equation}
We note that, in Eq.~\ref{eq:epsWLFE}, the local field effect (LFE) has been included~\cite{PhysRev.129.62}. Otherwise, the macroscopic dielectric function is 
\begin{equation}
\label{eq:epsWoLFE}
\epsilon_M(\mathbf q, \omega) =\epsilon_{00}(\mathbf q, \omega),
\end{equation}
and describe an inaccurate plasmon dispersion with a small red-shift, as shown in Fig.~\ref{fig:figure_add}.

For comparison and analysis, simpler approximation called independent particle (IP) is occupied. By setting $ K_{\mathbf G_1 \mathbf G_2}(\mathbf q) = 0$ in Eq.~\ref{eq:Dyson}, 
\begin{equation}
\chi_{\mathbf G \mathbf G^{\prime}}(\mathbf q, \omega) = \chi^0_{\mathbf G \mathbf G^{\prime}}(\mathbf q, \omega).
\end{equation}

\begin{figure}
\centering
\includegraphics[width=0.5\linewidth]{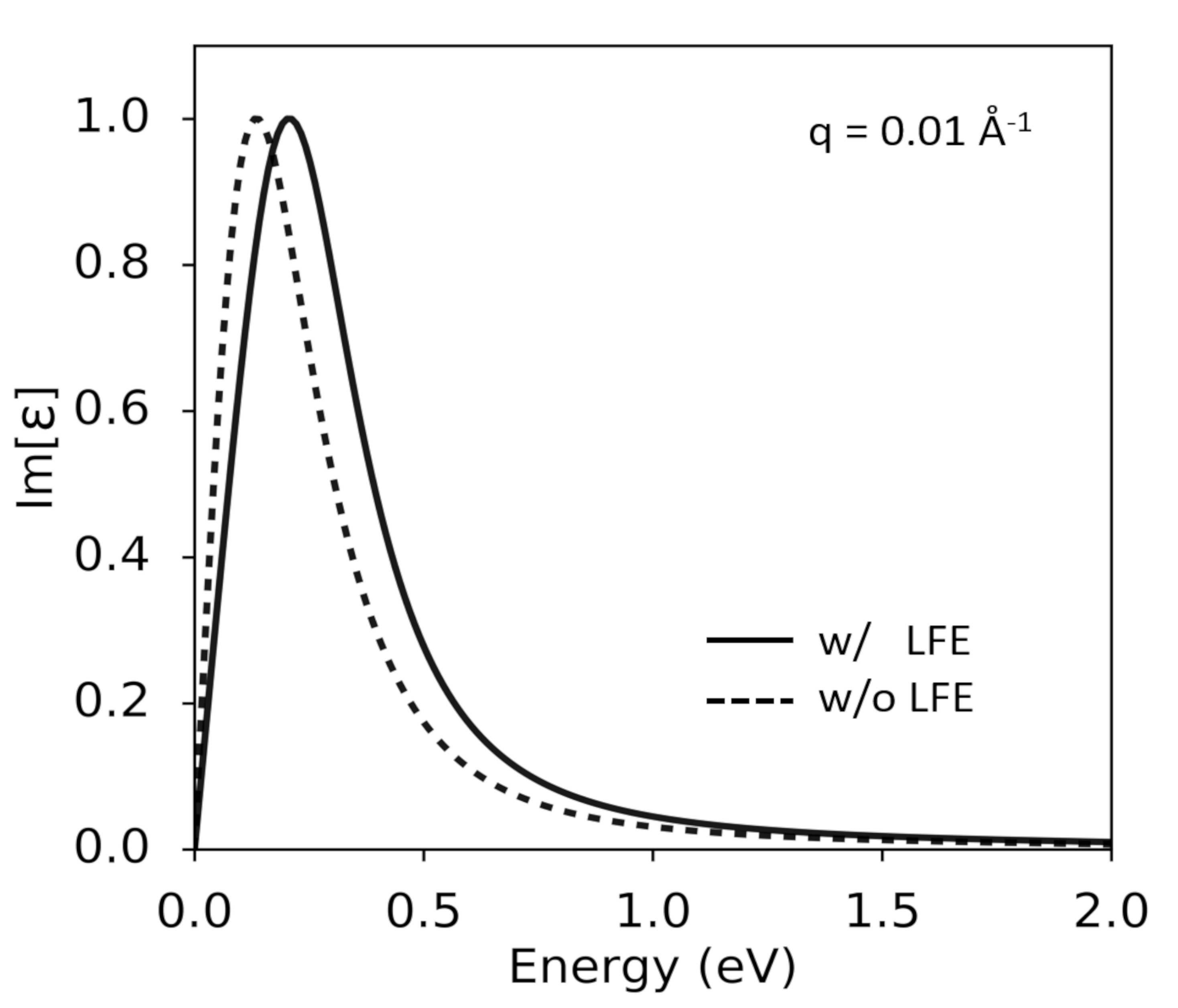}
\caption{Imaginary part of the dielectric function at long wavelength ($q=0.01~\mathrm{\AA}^{-1}$).}
\label{fig:sqepsilon}
\end{figure}
As shown in Fig.~\ref{fig:sqepsilon}, the optical response of the borophene, i.e. the long-wavelength plasmon response shows only one major peak at 0.27~eV. As q increases, the single branches split into two branches, as we discussed in the main text.

\begin{figure}
	\centering
	\includegraphics[width=1.0\linewidth]{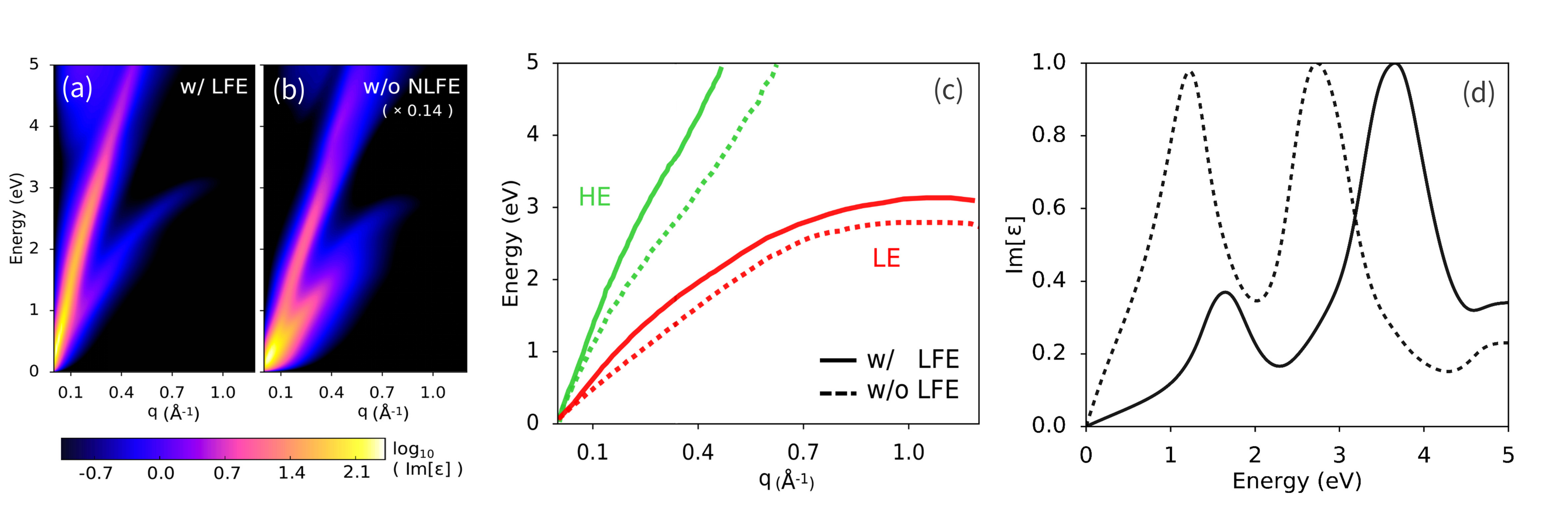}
	\caption{The comparison between imaginary part of the dielectric function with and without local field effect (a) (b) the contour plot of imaginary part of the dielectric function. (c) the peak positions of the plasmon in (a) and (b). (d) Imaginary part of the dielectric function as a function of energy at $q=0.35$~\AA$^{-1}$.}
	\label{fig:figure_add}
\end{figure}

\begin{figure}
\centering
\includegraphics[width=1.0\linewidth]{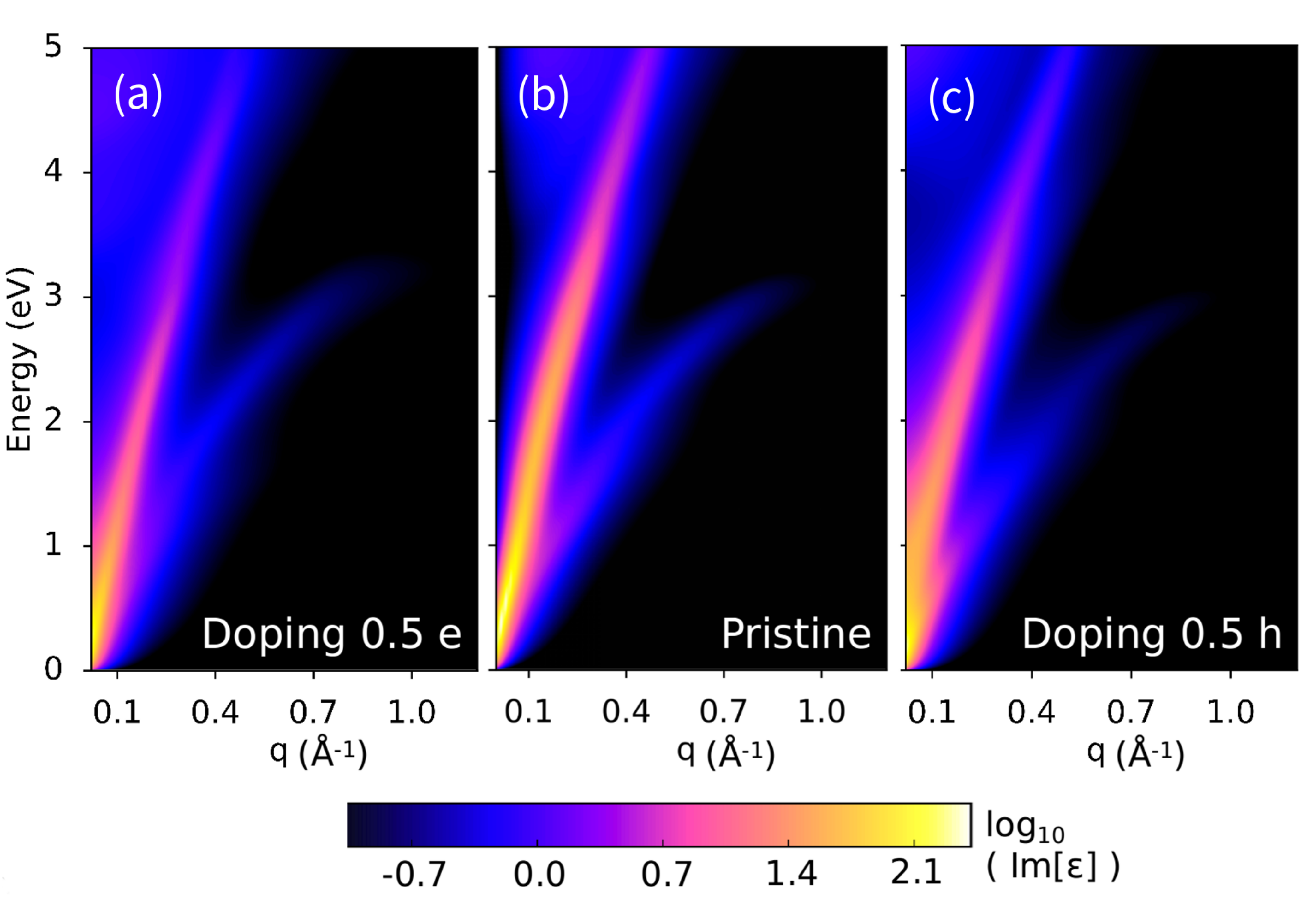}
\caption{Imaginary part of the dielectric function along $\Gamma$-X at different doping levels: (a) 0.5 e/(unit cell) doping, (b) without doping, and (c) 0.5 h/(unit cell) doping.}
\label{fig:DopingEELScontour}
\end{figure}

\begin{figure}
	\centering
	\includegraphics[width=1.0\linewidth]{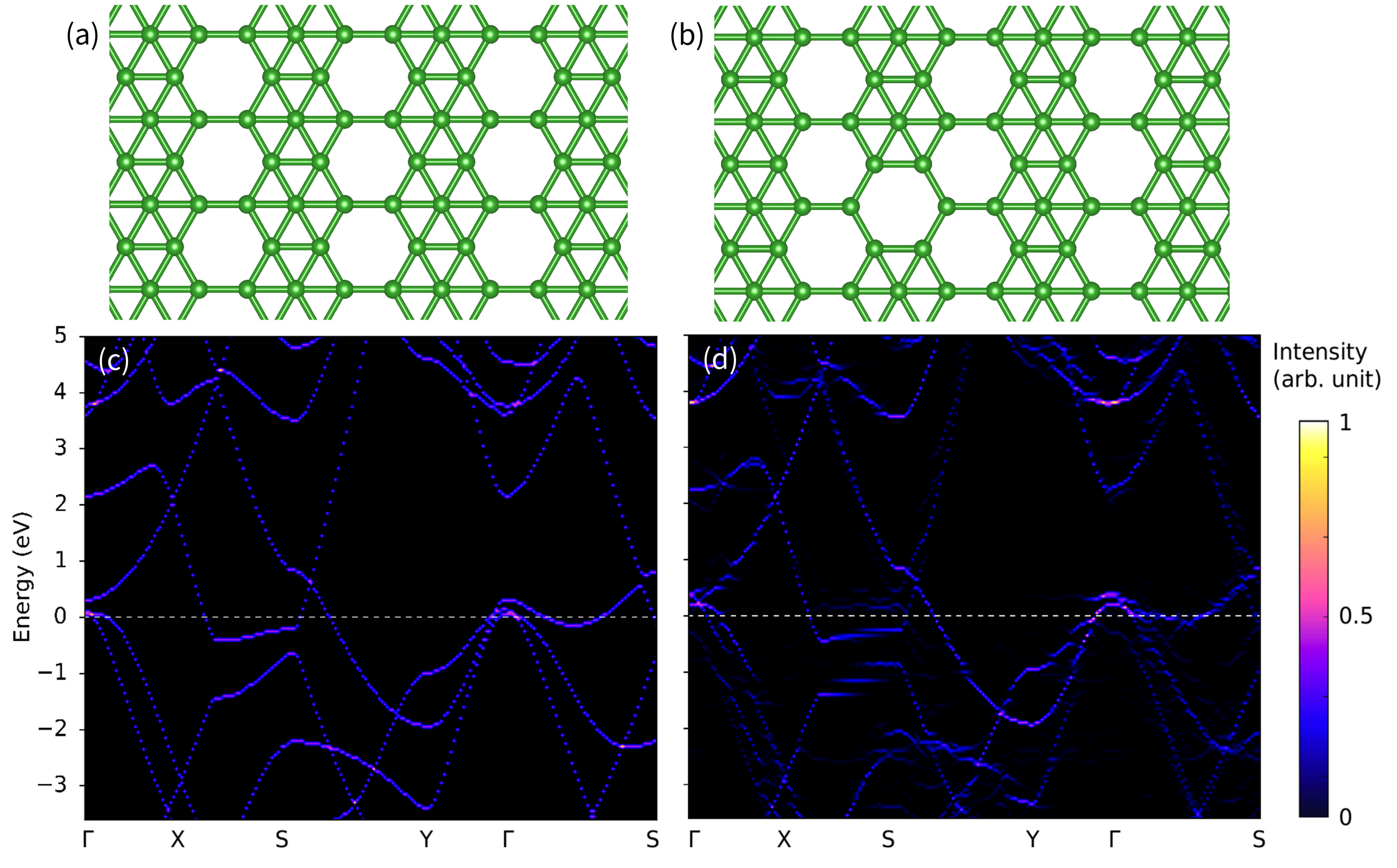}
	\caption{(a) Structure and (c) effective band structure (EBS) of pristine borophene with a $4\times 4$ super cell. (b) Structure and (d) EBS of borophene with a point defect in $4\times 4$ supercell. The EBS is calculated with modified BandUp package~\cite{Medeiros2014,Medeiros2015,Lian2017a}. }
	\label{fig:figureadd2}
\end{figure}

\end{document}